\documentclass[preprint,showpacs,preprintnumbers,amsmath,amssymb]{revtex4-2}

\usepackage{graphicx}
\usepackage{dcolumn}
\usepackage{bm}
\usepackage{color}

\begin{document}

\newif\ifplot
\plottrue
\newcommand{\RR}[1]{[#1]}
\newcommand{\intsum}{\sum \kern -15pt \int}
\newfont{\Yfont}{cmti10 scaled 2074}
\newcommand{\Y}{\hbox{{\Yfont y}\phantom.}}
\def\O{{\cal O}}
\newcommand{\bra}[1]{\left< #1 \right| }
\newcommand{\braa}[1]{\left. \left< #1 \right| \right| }
\def\Bra#1#2{{\mbox{\vphantom{$\left< #2 \right|$}}}_{#1}
\kern -2.5pt \left< #2 \right| }
\def\Braa#1#2{{\mbox{\vphantom{$\left< #2 \right|$}}}_{#1}
\kern -2.5pt \left. \left< #2 \right| \right| }
\newcommand{\ket}[1]{\left| #1 \right> }
\newcommand{\kett}[1]{\left| \left| #1 \right> \right.}
\newcommand{\scal}[2]{\left< #1 \left| \mbox{\vphantom{$\left< #1 #2 \right|$}}
\right. #2 \right> }
\def\Scal#1#2#3{{\mbox{\vphantom{$\left<#2#3\right|$}}}_{#1}
{\left< #2 \left| \mbox{\vphantom{$\left<#2#3\right|$}}
\right. #3 \right> }}

\title{Efficient emulator for solving three-nucleon continuum
  Faddeev equations with chiral three-nucleon force comprising any number
  of contact terms}

\author{H.~Wita{\l}a}
\email{henryk.witala@uj.edu.pl}
\affiliation{
M. Smoluchowski Institute of Physics, Jagiellonian University,
PL-30348 Krak\'ow, Poland}
\author{J.~Golak}
\affiliation{
M. Smoluchowski Institute of Physics, Jagiellonian University,
PL-30348 Krak\'ow, Poland}
\author{R.~Skibi\'nski}
\affiliation{
M. Smoluchowski Institute of Physics, Jagiellonian University,
PL-30348 Krak\'ow, Poland}

\date{Received: date / Accepted: date}

\begin{abstract}
  We demonstrate a computational scheme which drastically decreases the 
  required time to get  theoretical predictions
  based on chiral two- and three-nucleon forces  for 
  observables in three-nucleon continuum. For a 
  three-nucleon force containing N short-range terms  all
  workload is reduced to solving 
  N+1  Faddeev-type integral equations. That done, computation of
  observables for any combination of strengths of the contact terms is
  done in a flash.
  We  demonstrate on example of the elastic nucleon-deuteron scattering
  observables the high precision of the proposed emulator 
 and its capability to reproduce exact results.
\end{abstract}

\pacs{21.30.-x, 21.45.-v, 24.10.-i}

\maketitle

 Since the birth of nuclear physics the nuclear force problem
 has been at the centre of
 experimental and theoretical studies. Extensive efforts based 
 on purely phenomenological approaches or incorporating
 the meson-exchange picture have led to
 numerous nucleon-nucleon (NN) potentials, able to describe 
 a vast amount of available data \cite{mach_anp} with high precision. 
 In spite of the enormous progress
 in understanding properties of the two-nucleon interaction, applications of
 these ideas to the many-nucleon forces encountered consistency
 problems and called for a more systematic framework. A major
 breakthrough occurred with the emergence of the effective field theory (EFT)
 concept \cite{weinberg}, which paved the way
 for developing precise nuclear forces \cite{vankolck,epel_nn_n3lo,epel6a,machl6b}.

 The progress in constructing nuclear forces within the EFT approach
 is presently documented by the availability
 of  numerous high precision NN potentials. 
 Recently a new generation of chiral NN potentials
 was introduced and developed up to the fifth order (N$^4$LO) of
 chiral expansion
 by the
 Bochum-Bonn \cite{epel1,epel2} and Idaho-Salamanca \cite{entem2017} groups.
 These forces provide  a very
good description of the NN data set (Idaho-Salamanca) or the phase shifts and
mixing angles of the Nijmegen partial wave analysis \cite{nijmpwa}
(Bochum-Bonn), 
used to fix the low-energy constants accompanying the NN contact
interactions. 
The latest and most precise EFT-based NN interaction is the  semilocal
momentum-space (SMS)
regularized chiral potential of the Bochum group \cite{preinert},
 developed up to N$^4$LO and even including some terms from the next order
 of chiral expansion (N$^4$LO$^+$).
 In this potential a new momentum-space regularization scheme
 has been employed 
 for the long-range contributions 
 and a nonlocal
 Gaussian regulator has been applied to the minimal set of independent
 contact interactions. 
 This new approach can be straightforwardly utilised to regularize
 also three-nucleon (3N) forces. 
 That new family of semilocal chiral potentials provides an outstanding
 description of the NN data.

 Applications of the EFT approach in the form of chiral
 perturbation theory (ChPT) have resulted not only in the theoretically
 well grounded
 NN potentials but also for the first time have given a
 possibility to apply in practical calculations NN forces augmented by 
 consistent 3N interactions, derived within the same  formalism.
 Understanding of nuclear spectra and reactions based on these 
  consistent chiral two- and many-body forces has become a hot topic
  of present day few-body studies \cite{epel2019}.

  The first nonvanishing contributions to the 3N force (3NF) appear
  at next-to-next-to-leading order of chiral expansion (N$^2$LO)
 \cite{vankolck,epel2002} and comprise
in addition to the $2\pi$-exchange term two contact contributions with strength
parameters $c_D$ and $c_E$ \cite{epel_tower}. 
 The difficult task to derive the chiral
3NF at N$^3$LO  
was accomplished in \cite{3nf_n3lo_long,3nf_n3lo_short}.
 At that order five different topologies contribute
 to 3NF. Three of them are of long-range character \cite{3nf_n3lo_long}
 and are given by two-pion ($2\pi$) exchange graphs, by
 two-pion-one-pion ($2\pi-1\pi$) exchange graphs, and
by the  ring diagrams. They are supplemented by the short-range
two-pion-exchange-contact ($2\pi$-contact) term and by the leading
relativistic corrections to 3NF \cite{3nf_n3lo_short}.
 The 3NF at N$^3$LO order
does not involve any new unknown low-energy constants (LECs) and depends only
on two parameters, $c_D$ and $c_E$ that parameterize the leading
one-pion-contact term and the 3N contact term present already at N$^2$LO.
The $c_D$ and $c_E$ values need to be then fixed at this order,
as at N$^2$LO, from a fit to few-nucleon data.
At the higher order, N$^4$LO, in addition to long- and
intermediate-range interactions generated by pion-exchange
diagrams \cite{krebs2012,krebs2013}, the chiral N$^4$LO 3NF involves thirteen 
purely short-range
operators, which have been worked out in \cite{girlanda2011}.

Since the advent of numerically exact three-nucleon continuum Faddeev 
calculations the elastic nucleon-deuteron (Nd)
 scattering and the deuteron breakup reaction  
 have been a powerful tool to test  modern models of
 the nuclear forces \cite{glo96,pisa,hanover}  
  and  the question
 about the importance of 3NF has developed into the main topic of 
 3N system  studies. That issue has been given a new impetus 
 by the ChPT-based achievements
 and the possibility to apply consistent
 two- and many-body nuclear forces, derived within this framework,
 in 3N continuum calculations.

 Using chiral 3NF in 3N continuum 
 requires numerous time consuming computations  with varying strengths
 of the contact terms in order to establish their values.  
 They can be determined for example
 from the $^3$H binding energy and the minimum 
 of the elastic Nd scattering  
 differential cross section at the energy ($E_{lab}\approx 70$~MeV),
 where the effects of 3NF start to emerge in elastic Nd scattering
 \cite{wit2001,wit98}.  
Specifically at N$^2$LO, after establishing the so-called ($c_D,c_E$)
correlation line,
which for a particular chiral NN potential combined
with a N$^2$LO 3NF gives pairs of ($c_D,c_E$) values reproducing 
the $^3$H binding energy, 
a fit to  experimental data for the elastic Nd cross section is performed to
determine the $c_D$ and $c_E$ strengths.
Fine-tuning of the 3N Hamiltonian parameters requires an extensive
analysis of available 3N elastic Nd scattering and breakup data. That
ambitious goal calls for a significant reduction of computer time necessary to 
solve the 3N Faddeev equations and to calculate the observables.
Thus finding an efficient emulator for exact
solutions of the 3N Faddeev equations seems to be essential and of
high priority.

In Ref.~\cite{pert} we proposed such an emulator
  which enables us to reduce
  significantly the required time of calculations. We 
 tested its efficiency 
as well as ability to accurately reproduce exact solutions of 3N Faddeev
equations. 
In the present study we introduce a new computational
 scheme, based on the perturbative approach
 of \cite{pert}, which even by far more
  reduces the computer time required to
obtain the observables in the elastic nucleon-deuteron scattering
and deuteron breakup reactions at any energy, and which is well-suited for
 calculations with varying strengths of the contact terms
 in a chiral 3NF.
 Before presenting this new emulator, for the reader's convenience
 we shortly outline the main points of the 3N Faddeev formalism
 and of the perturbative treatment of Ref.~\cite{pert}. 
 For details of the formalism  and numerical performance 
  we refer to ~\cite{glo96,wit88,hub97,book}.

Neutron-deuteron (nd) scattering with nucleons interacting
via NN interactions $v_{NN}$ and a 3NF $V_{123}=V^{(1)}+V^{(2)}+V^{(3)}$, is
described in terms of a breakup operator $T$ satisfying the
Faddeev-type integral equation~\cite{glo96,wit88,hub97}
\begin{eqnarray}
T\vert \phi \rangle  &=& t P \vert \phi \rangle +
(1+tG_0)V^{(1)}(1+P)\vert \phi \rangle + t P G_0 T \vert \phi \rangle \cr 
&+& 
(1+tG_0)V^{(1)}(1+P)G_0T \vert \phi \rangle \, .
\label{eq1a}
\end{eqnarray}
The 2N $t$-matrix $t$ is the solution of the
Lippmann-Schwinger equation with the interaction
$v_{NN}$.   $V^{(1)}$ is the part of a 3NF which is 
symmetric under the interchange of nucleons $2$ and $3$: $V_{123}=V^{(1)}(1+P)$.
 The permutation operator $P=P_{12}P_{23} +
P_{13}P_{23}$ is given in terms of the transposition operators,
$P_{ij}$, which interchange nucleons $i$ and $j$.  The initial state 
$\vert \phi \rangle = \vert \vec {q}_0 \rangle \vert \phi_d \rangle$
describes the free motion of the neutron and the deuteron 
  with the relative momentum
  $\vec {q}_0$  and contains the internal deuteron wave function
  $\vert \phi_d \rangle$.
  $G_0$ is the free three-body resolvent.
The amplitude for elastic scattering leading to the 
 final nd state $\vert \phi ' \rangle$ is then given by~\cite{glo96,hub97}
\begin{eqnarray}
\langle \phi' \vert U \vert \phi \rangle &=& \langle \phi' 
\vert PG_0^{-1} \vert 
\phi \rangle  
 + \langle 
\phi'\vert  V^{(1)}(1+P)\vert \phi \rangle  \cr
&+& \langle \phi' \vert V^{(1)}(1+P)G_0T\vert  \phi \rangle +
\langle \phi' \vert PT \vert \phi \rangle ~,
\label{eq3}
\end{eqnarray}
while the  amplitude for the breakup reaction reads
\begin{eqnarray}
\langle  \vec p \vec q \vert U_0 \vert \phi \rangle &=&\langle 
 \vec p \vec q \vert  (1 + P)T\vert
 \phi \rangle ,
\label{eq3_br}
\end{eqnarray}
where the free  breakup channel state $\vert  \vec p \vec q \rangle $
is defined in terms of the  Jacobi (relative) momenta $\vec p$
and $\vec q$. 

We solve Eq.~(\ref{eq1a}) in the momentum-space partial-wave basis
$\vert p q \alpha \rangle$, determined by 
the magnitudes of the 
Jacobi momenta $p$ and $q$ and a set of discrete quantum numbers $\alpha$
comprising 
the 2N subsystem spin,
orbital and total angular momenta $s, l$ and $j$, as well as 
the spectator nucleon orbital
and total angular momenta with respect to the center of mass (c.m.) of the 2N
subsystem, $\lambda$ and $I$:
\begin{eqnarray}
\vert p q \alpha \rangle \equiv \vert p q (ls)j (\lambda \frac {1} {2})I (jI)J
  (t \frac {1} {2})T \rangle ~.
\label{eq4a}
\end{eqnarray}
The total 2N and spectator angular momenta $j$ and $I$ as well as isospins
$t$ and $\frac {1} {2}$, are finally 
coupled to the total angular momentum $J$ and isospin $T$ of the 3N system.
In practice a converged solution of Eq.~(\ref{eq1a})
using partial wave decomposition
in momentum space at a given energy $E$ requires taking all 3N partial wave
states up to the 2N angular momentum $j_{max}=5$ 
and the 3N angular momentum $J_{max}=\frac{25}{2}$, with
the 3N force acting up to the 3N total
angular momentum $J=7/2$. The number of resulting
partial waves (equal to the number of coupled integral equations in
two continuous
variables $p$ and $q$)
amounts to $142$. The required computer time to get one solution on a 
personal computer is about
$\approx 2$~h. In the case when such calculations have to be performed for a
big number of varying 3NF parameters, time restrictions
become prohibitive. Fortunately, the perturbative approach of  Ref.~\cite{pert}
leads to a significant reduction of the required computational time.

Let us consider a chiral 3NF at a given order of chiral expansion with
variable strengths of its contact  terms.
 The 3NF at N$^2$LO has one parameter-free term (2$\pi$-exchange
 contribution) and two short-range terms with strength
 parameters $c_D$ and $c_E$.
 At N$^3$LO there are more contributing  parameter-free parts 
  but again only two contact terms.
  At N$^4$LO parameter-free contributions are supplemented by
  fifteen short-range 
 terms with strengths: $c_D$, $c_E$,  $c_{E_1}$, ..., $c_{E_{13}}$.
 All these contact terms are restricted to 
 small 3N total angular momenta and
 to only few partial wave states for a given total 3N angular momentum 
 $J$ and parity $\pi$. For example for $J^{\pi}=7/2^{\pm}$ all
 matrix elements $< p q \alpha \vert V^{(1)}  \vert p' q' \alpha' >$
 proportional to 
 $c_{E_1}$ and $c_{E_7}$  vanish, while the $c_D$ and $c_E$ terms are nonzero only
 for a restricted number of  $\alpha, \alpha'$ pairs  
 (mostly these containing $^1S_0$ and
   $^3S_1-^3D_1$ quantum numbers) \cite{epel2002,epel_tower}.  
   Bearing that in mind and taking into account the fact that contact 
   terms yield a small
   contribution to the 3N potential energy compared to the
   leading, parameter-free part, it is possible to apply a perturbative
 approach in order to include the contact terms.

 We split the $V^{(1)}$ part of a
 3NF into a parameter-free term $V(\theta_0)$
 and a sum of $N$ contact terms $c_i \Delta V_i$ 
   with strengths $c_i$: 
\begin{eqnarray}
  V^{(1)}  &=&V(\theta_0)   +   \Delta V(\theta)
  = V(\theta_0)   +  \sum_{i=1}^N c_i \Delta V_i  ~,
\label{eq4}
\end{eqnarray}
with $\theta_0=(c_i=0,i=1,\dots,N)$ and $\theta=(c_i,i=1,\dots,N)$ being 
the sets
of contact terms strength values, 
 for which we would like to find solution of
 Eq.~(\ref{eq1a}).

We divide the 3N partial wave states into two sets: $\beta$ and
the remaining one, $\alpha$. The $\beta$ set is defined by nonvanishing
matrix elements
of $\Delta V(\theta)$.
 Introducing $T(\theta_0)$ and $\Delta T(\theta)$ such that 
$T \equiv T(\theta)=T(\theta_0) + \Delta T(\theta)$,
 and using the fact, that $\Delta V(\theta)$ has nonvanishing elements only
 for channels $\vert \beta \rangle$, one gets from Eq.~(\ref{eq1a})
 (omitting the Jacobi
momenta in notation of partial wave states) two separate equations for
 $\langle \alpha \vert T (\theta_0) \vert \phi \rangle$ and
 $\langle \alpha \vert \Delta T (\theta) \vert \phi \rangle$ \cite{pert}:
\begin{eqnarray}
  \langle \alpha \vert T (\theta_0) \vert \phi \rangle
  &=&
 \langle \alpha \vert t P \vert \phi \rangle +
 \langle \alpha \vert (1+tG_0)  V(\theta_0) (1+P)\vert \phi  \rangle \cr 
 &+& \langle \alpha \vert t P G_0 T(\theta_0) \vert \phi \rangle \cr
 &+& \langle \alpha \vert (1+tG_0) V (\theta_0) (1+P)G_0T(\theta_0)
  \vert \phi \rangle  \cr
  \langle \alpha \vert \Delta T (\theta) \vert \phi \rangle &=&
   \langle \alpha \vert t P G_0 \Delta T(\theta) \vert \phi \rangle \cr
   + \langle \alpha \vert (1&+&tG_0) V (\theta_0) (1+P)G_0 \Delta T(\theta)
  \vert \phi \rangle ~,
\label{eq7}
\end{eqnarray}
as well as  for   $\langle \beta \vert T (\theta_0) \vert \phi \rangle$ and
 $\langle \beta \vert \Delta T (\theta) \vert \phi \rangle$:
\begin{eqnarray}
  \langle \beta \vert T (\theta_0) \vert \phi \rangle
  &=&
 \langle \beta \vert t P \vert \phi \rangle +
 \langle \beta \vert (1+tG_0)  V(\theta_0) (1+P)\vert \phi  \rangle \cr
 &+& \langle \beta \vert t P G_0 T(\theta_0) \vert \phi \rangle \cr
 &+& \langle \beta \vert (1+tG_0) V (\theta_0) (1+P)G_0T(\theta_0)
  \vert \phi \rangle  \cr
  \langle \beta \vert \Delta T (\theta) \vert \phi \rangle &=&
 \langle \beta \vert (1+tG_0) \Delta V(\theta) (1+P)\vert \phi  \rangle \cr
 &+& \langle \beta \vert (1+tG_0) \Delta V(\theta) (1+P)G_0 T(\theta_0)
 \vert \phi  \rangle \cr
 &+& \langle \beta \vert (1+tG_0) V (\theta_0) 
 (1+P)G_0 \Delta T(\theta)   \vert \phi \rangle  \cr
 &+& \langle \beta \vert (1+tG_0) \Delta V(\theta)
 (1+P)G_0 \Delta T(\theta)   \vert \phi \rangle  \cr 
 &+& \langle \beta \vert t P G_0 \Delta T(\theta) \vert \phi \rangle ~.
\label{eq8}
\end{eqnarray}

The first equations in (\ref{eq7}) and (\ref{eq8}) are the
Faddeev equations (\ref{eq1a}) for $T(\theta_0)$. Since the two leading terms
for $\langle \beta \vert \Delta T (\theta) \vert \phi \rangle$ in (\ref{eq8})
 are of the order of $\Delta V(\theta)$ then
$\langle \alpha \vert \Delta T (\theta) \vert \phi \rangle \approx 0$ and
 the second equation in the set (\ref{eq8}) for
 $\langle \beta \vert \Delta T (\theta) \vert \phi \rangle$ can be solved
 within the set of channels $\vert \beta \rangle$ only.  
 Using this solution,
 $\langle \alpha \vert \Delta T (\theta) \vert \phi \rangle$ is then 
computed  by:
\begin{eqnarray}
\langle \alpha \vert \Delta T (\theta) \vert \phi \rangle &=&
\langle \alpha \vert t P G_0  \sum_{\beta} \int_{p'q'}
\vert p' q' \beta \rangle \langle p' q' \beta \vert
\Delta T(\theta)
 \vert \phi \rangle \cr 
&+& \langle \alpha \vert (1+tG_0) V (\theta_0) (1+P)G_0 \cr
&&\sum_{\beta} \int_{p'q'}  \vert p'q' \beta \rangle
\langle p' q' \beta \vert   \Delta T(\theta)
  \vert \phi \rangle ~.
\label{eq9}
\end{eqnarray}

Finally, $T(\theta)$ is calculated as
\begin{eqnarray}
  \langle \alpha \vert T (\theta) \vert \phi \rangle &=&
  \langle \alpha \vert T (\theta_0) \vert \phi \rangle
  +  \langle \alpha \vert \Delta T (\theta) \vert \phi \rangle  \cr
  \langle \beta \vert T (\theta) \vert \phi \rangle &=&
  \langle \beta \vert T (\theta_0) \vert \phi \rangle +
  \langle \beta \vert \Delta T (\theta) \vert \phi \rangle ~.
\label{eq10}
\end{eqnarray}

The outlined above procedure constitutes the perturbative approach
 of Ref.~\cite{pert}. In short, one solves
the 3N Faddeev equation (\ref{eq1a}) exactly with  
 the  NN potential combined with the 3NF restricted  to
the parameter free term  
(set $\theta_0=(0,\dots,0)$).
 The solution with the $\theta_0$ set forms a starting point in the 
perturbative treatment  of Eqs.~(\ref{eq7})-(\ref{eq10}) and has to be
calculated only once, regardless of how many variations of strength parameters
are required. 
In the next step, the proper  perturbative  treatment is performed,
solving first the second
equation in set (\ref{eq8}). 
Having determined $\langle \alpha \vert \Delta T (\theta) \vert \phi \rangle$
from  Eq.(\ref{eq9}) the emulator solution of Eq.~(\ref{eq10}) is calculated 
(in the following this emulator  will be denoted by $E\Delta T$).  
 That  allows one to reduce
the required computation time and to reproduce surprisingly well
the exact predictions for neutron-deuteron (nd) elastic scattering 
as well as for nd breakup observables \cite{pert}. 
 To be specific, taking set $\vert \beta \rangle$ which includes  all 
  2N states with the total 2N angular momenta $ j\le 2$,  
  leads to a reduction of the         
 computing time by a factor of approximately $4$ 
 in comparison to the exact calculations.
 Note that it takes approximately $30$~minutes on a personal
 computer to solve Eq.~(\ref{eq1a}), provided that the 
 $V(\theta_0) (1+P)$ and $V(\theta_i) (1+P)$ 
 kernels, acting in 
  $ (1+tG_0) V (\theta) (1+P)G_0T(\theta)
  \vert \phi \rangle$ term of Eq.~(\ref{eq1a}), 
  are prepared in advance, with the strengths
  $\theta_i=(c_i=1, c_{k\ne i}=0)$. 

In spite of such a large reduction, the computational time can be
even further decreased and calculation of 3N continuum observables
 made in a flash.
  This notion is based on the observation that among three kernel-terms  
  in the second equation of set (\ref{eq8}), it is possible 
  (because of the smallness of $\Delta V(\theta)$)
  to neglect the term 
 $ \langle \beta \vert (1+tG_0) \Delta V(\theta)
 (1+P)G_0 \Delta T(\theta)   \vert \phi \rangle $.
  The resulting integral equation for
  $\langle \beta \vert \Delta T (\theta) \vert \phi \rangle$: 
\begin{eqnarray}
  \langle \beta \vert \Delta T (\theta) \vert \phi \rangle &=&
 \langle \beta \vert (1+tG_0) \Delta V(\theta) (1+P)\vert \phi  \rangle \cr
 &+& \langle \beta \vert (1+tG_0) \Delta V(\theta) (1+P)G_0 T(\theta_0)
 \vert \phi  \rangle \cr
 &+& \langle \beta \vert (1+tG_0) V (\theta_0) 
 (1+P)G_0 \Delta T(\theta)   \vert \phi \rangle  \cr
 &+& \langle \beta \vert t P G_0 \Delta T(\theta) \vert \phi \rangle ~,
\label{eq11}
\end{eqnarray}
 permits one to transfer the linear dependence  on the strengths $c_i$ from 
 the $\Delta V(\theta)$ on  
 the  $\Delta T(\theta)$. Namely, let
 $ \langle \beta \vert \Delta T_i  \vert \phi \rangle $ be a solution
 of Eq.(\ref{eq11}) for a
 set $\theta_i=(c_i=1, c_{k \ne i}=0)$:
 \begin{eqnarray}
  \langle \beta \vert \Delta T_i \vert \phi \rangle &\equiv&
 \langle \beta \vert (1+tG_0) \Delta V_i (1+P)\vert \phi  \rangle \cr
 &+& \langle \beta \vert (1+tG_0) \Delta V_i (1+P)G_0 T(\theta_0)
 \vert \phi  \rangle \cr
 &+& \langle \beta \vert (1+tG_0) V (\theta_0) 
 (1+P)G_0 \Delta T_i   \vert \phi \rangle  \cr
 &+& \langle \beta \vert t P G_0 \Delta T_i \vert \phi \rangle ~.
\label{eq12}
\end{eqnarray}
Multiplying (\ref{eq12}) by $c_i$ and
 summing over $i$ one gets:
 \begin{eqnarray}
  \langle \beta \vert \sum_i c_i \Delta T_i \vert \phi \rangle &\equiv&
 \langle \beta \vert (1+tG_0) \sum_i c_i \Delta V_i (1+P)\vert \phi  \rangle \cr
 + \langle \beta \vert (1&+&tG_0) \sum_i c_i \Delta V_i (1+P)G_0 T(\theta_0)
 \vert \phi  \rangle \cr
 + \langle \beta \vert (1&+&tG_0) V (\theta_0) 
 (1+P)G_0 \sum_i c_i \Delta T_i   \vert \phi \rangle  \cr
 &+& \langle \beta \vert t P G_0 \sum_i c_i \Delta T_i \vert \phi \rangle ~,
\label{eq13}
\end{eqnarray}
 and the solution of Eq.~(\ref{eq11}) is given by:
\begin{eqnarray}
  \langle \beta \vert  \Delta T(\theta) \vert \phi \rangle &=&
  \sum_{i=1}^N c_i  \langle \beta \vert  \Delta T_i  \vert \phi \rangle ~.
\label{eq14}
\end{eqnarray}

In this way at a given  energy the computation of  observables in the elastic
Nd scattering and deuteron breakup reaction for  any combination of strengths
$c_i$ of contact terms is reduced to solving  once  $N+1$  
  Faddeev equations: one equation for $T(\theta_0)$
 and N equations for $\Delta T_i$. In the first step,
 solution for
 $ \langle \alpha  (\beta) \vert T (\theta_0) \vert \phi \rangle $  is found. 
 Then  Eq.~(\ref{eq12}) is solved for
$ \langle \beta \vert \Delta T_i \vert \phi \rangle $, from which  the 
$ \langle \alpha \vert T_i  \vert \phi \rangle $ is calculated by:
\begin{eqnarray}
\langle \alpha \vert \Delta T_i \vert \phi \rangle &=&
\langle \alpha \vert t P G_0  \sum_{\beta} \int_{p'q'}
\vert p' q' \beta \rangle \langle p' q' \beta \vert
\Delta T_i \vert \phi \rangle \cr 
&+& \langle \alpha \vert (1+tG_0) V (\theta_0) (1+P)G_0 \cr
&&\sum_{\beta} \int_{p'q'}  \vert p'q' \beta \rangle
\langle p' q' \beta \vert   \Delta T_i  \vert \phi \rangle ~.
\label{eq15}
\end{eqnarray}

The above computations need to be done only once and then for any combination
of the strengths $c_i$
$\langle \alpha ( \beta) \vert T(\theta=(c_i,i=1,\dots,N) ) \vert \phi \rangle $
is obtained by trivial summation:
\begin{eqnarray}
  \langle \alpha \vert T (\theta) \vert \phi \rangle &=&
  \langle \alpha \vert T (\theta_0) \vert \phi \rangle
  + \sum_i c_i \langle \alpha \vert \Delta T_i \vert \phi \rangle  \cr
  \langle \beta \vert T (\theta) \vert \phi \rangle &=&
  \langle \beta \vert T (\theta_0) \vert \phi \rangle + \sum_i c_i
  \langle \beta \vert \Delta T_i \vert \phi \rangle ~.
\label{eq16}
\end{eqnarray}

For a calculation of elastic scattering observables 
the required sum of the second and the  third  term in Eq.~(\ref{eq3})
is obtained by:
\begin{eqnarray}
 && \langle  \alpha \vert  V^{(1)}(\theta)(1+P)\vert \phi \rangle
  + \langle  \alpha \vert V^{(1)}(\theta)(1+P)G_0T(\theta)
  \vert  \phi \rangle = \cr
 && \langle  \alpha \vert  V(\theta_0)(1+P)\vert \phi \rangle
  +  \langle  \alpha \vert  V(\theta_0)(1+P)G_0 T(\theta_0)
  \vert \phi \rangle \cr
&&+ \sum_i c_i [ \langle  \alpha \vert \Delta V_i(1+P) \vert \phi \rangle
    +  \langle  \alpha \vert \Delta V_i(1+P)G_0 T(\theta_0)
    \vert \phi \rangle \cr
&&    +  \langle  \alpha \vert V(\theta_0) (1+P)G_0 \Delta T_i 
  \vert \phi \rangle ] \cr  
&& + \sum_{i,k} c_i c_k  \langle  \alpha \vert \Delta V_i (1+P)G_0 \Delta T_k
  \vert \phi \rangle ~.
\label{eq17}
\end{eqnarray}

The above computational scheme forms the new emulator, which will be denoted
in the following by $E\Delta T_i$.
To check its  quality and efficiency as well as to compare it
with $E\Delta T$ we have chosen, 
 as in Ref.~\cite{pert}, the SMS N$^4$LO$^+$ chiral potential of the
Bochum group \cite{preinert}, with  the regularization
 cutoff $\Lambda = 450$~MeV, and combined it with the chiral N$^2$LO
 3NF.
 We solved the 3N Faddeev equation (\ref{eq1a}) exactly at two incoming
 neutron energies 
$E=70$ and $190$~MeV with that combination 
as well as with this NN potential supplemented with the 
parameter free $2\pi$-exchange term of the N$^2$LO 3NF 
(set $\theta_0=(c_D=0, c_E=0)$).
The first energy was taken from a region
where 3NF effects start to appear in 3N continuum observables
and the second one from a range with well-developed 3NF
effects \cite{wit2001,wit98}. The
solution with the $\theta_0$ set together with solutions
 $\langle \beta \vert \Delta T_i \vert \phi \rangle$
of  N integral equations (\ref{eq12}) with the set of 3N
 channels $ \vert \beta \rangle$,  
 comprising all 2N states with the total angular momentum $ j\le 2$, 
 form a starting point in the proposed
computational scheme and have to be
calculated only once.
 Also
 $\langle \alpha \vert \Delta T_i (\theta) \vert \phi \rangle$
 is  computed only once according to Eq.(\ref{eq15}), and, in
  parallel with those
 calculations, all terms in Eq.(\ref{eq17}) independent from $c_i$.
 Based on these quantities
 we calculated the emulator solution of Eq.~(\ref{eq16}).  
 Our new scheme performed for $N=2$ leads to 
 reduction by a factor of $\approx 50$ of the 
 computer time required by  the perturbative approach of Ref.~\cite{pert}.
 Computation  of all elastic scattering observables or observables in one
  exclusive breakup
 geometry for a particular set
 of strengths $\theta=(c_i, i=1,\dots,N)$,  
 in the perturbative approach requires $\approx 8$~minutes
 of a personal  computer time, while the new scheme requires only
 $\approx 10$~seconds. 
  Since the calculation of 
 $\langle \alpha \vert T (\theta) \vert \phi \rangle$ and
 $ \langle \beta \vert T (\theta) \vert \phi \rangle$ according
 to Eq.~(\ref{eq16}) takes practically no time, the
 drastic  reduction of time in the new scheme is independent
 from the number of the contact terms $N$.

\begin{figure}    
\includegraphics[scale=0.4,clip=true]{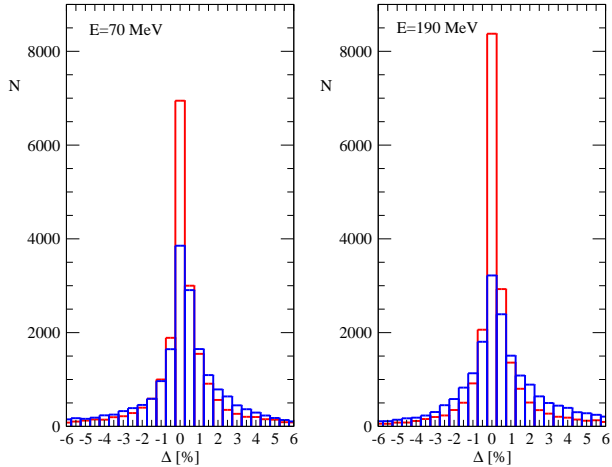}
\caption{
  (color online)
  Histograms of percentage deviations from exact results
  of the predictions by the emulator $E\Delta T$ 
  (red lines) or  $E\Delta T_i$  (blue lines),  
  for $51$ nonvanishing observables of the
  elastic nd scattering and 6 sets of $(c_D,c_E)$ strengths.
  For details see text.
}
\label{fig1}
\end{figure}

 The emulators $E \Delta T_i$ and $E \Delta T$
 are approximations of the exact results. 
 To estimate the quality of these approximations 
 we show in Fig.~\ref{fig1} at both energies
 the  distributions of percentage deviations 
 from exact results $O_{exact}$  predictions of both emulators $O_{appr}$:  
 $\Delta= {(O_{appr}-O_{exact})} / {O_{exact}} \times 100.0$, 
 for all nonvanishing nd elastic scattering observables $O$ 
  (cross section,  nucleon and deuteron analyzing powers as well
 as spin correlation and spin transfer coefficients between all participating
 particles, altogether 51 observables). They were calculated
 on a uniform grid
 of $73$ c.m. angles $\theta_{c.m.} \in (0^o,180^o)$. 
 These distributions encompass
 six sets of strengths values $(c_D,c_E)=(1, 1), (2, 1), (4, 1)$, 
 $(6, 1)$, $(8, 1)$, and  $(10, 1)$
 (according to the notation of Refs.~\cite{epel2002,epel_tower}).
 The precision of both emulators at both energies is similar and amounts
 to $\approx 1-2$~\%, with $E \Delta T$ being slightly more
 precise than $E \Delta T_i$.

\begin{figure}    
\includegraphics* [109,96] [332,277] {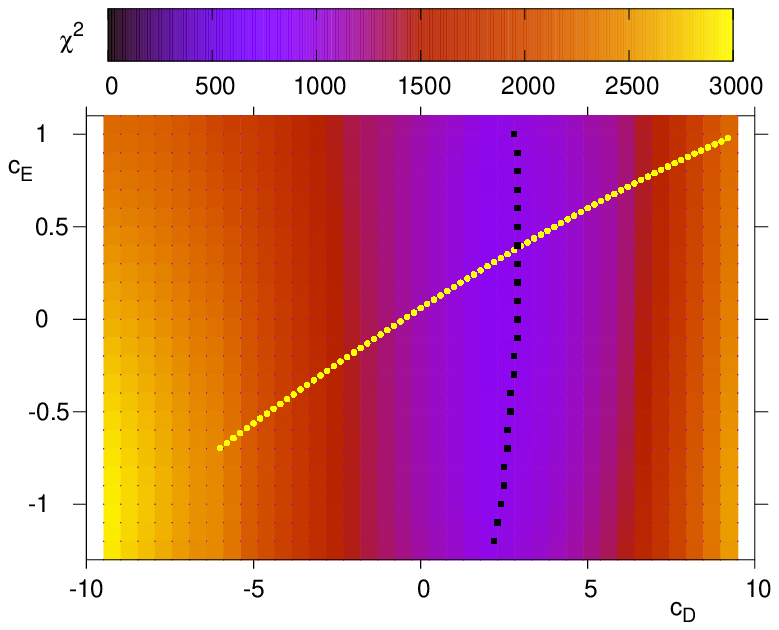}  
\caption{
  (color online)
  The color map of $\chi^2$ values for nd  elastic scattering cross section
  at $E=70$~MeV, in the plane of strengths ($c_D,c_E$) of
  N$^2$LO 3NF contact terms. It was obtained with the $E\Delta T_i$  emulator
 for combination
  of chiral NN  N$^4$LO$^+$ potential and N$^2$LO 3NF and computed from 
   about $5000$ strengths combinations.
  The proton-deuteron cross section data were taken
  from \cite{sekigu2002} and $\chi^2$ calculated for c.m. angles in the
  range $\theta_{c.m.} \in (62.18^o,158.33^o)$.
 The (black) squares show the position of the
  $\chi^2$-minimum at given $c_E$-value and (yellow) dots depict
  the $^3$H correlation line.
}
\label{fig2}
\end{figure}

\begin{figure}    
\includegraphics[scale=0.5,clip=true]{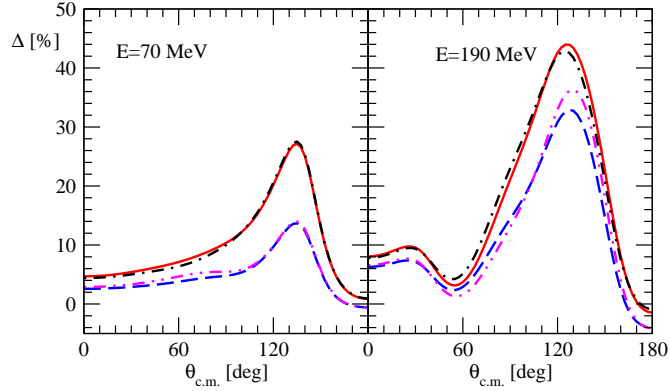}
\caption{
  The (red) solid lines show  percentage changes of the nd
  elastic scattering cross
  section at $E=70$ and $190$~MeV, predicted with SMS N$^4$LO$^+$ NN potential,
  induced  by 
  N$^2$LO 3NF with strengths of the contact terms $c_D=2.910$ and $c_E=0.385$.
  The (blue) dashed lines correspond to changes caused by a 3NF
  parameter-free term, 
  while the (black) dash-dotted and (magenta) dash-double-dotted curves to
  changes by a 3NF parameter free + D and parameter free + E terms,
  respectively.
}
\label{fig3}
\end{figure}

To demonstrate the power of the new emulator we show
in Fig.~\ref{fig2} the color map of $\chi^2$ values in the plane
 of strengths ($c_D,c_E$) for elastic scattering
cross section at $E=70$~MeV. That map
was obtained with about $5000$  values of strength combinations, both
varied in step of $0.1$. The minimum of $\chi^2$ valley is indicated 
by  black squares. It crosses with $\chi^2$ values for
 the $^3$H correlation line, shown by yellow
dots, in their minimum. The resulting values of strengths and their errors 
are $c_D=2.910 \pm 0.140$
and $c_E=0.385 \pm 0.015$. Using these strengths  we demonstrate in Fig.~\ref{fig3}
importance of contributions to the cross section from individual terms  
of the N$^2$LO 3NF. We show percentage
changes in the cross section values calculated solely with the NN potential 
by adding the parameter-free
term itself, 
and by the combination of that term with the D- or E-terms.   
 The 2$\pi$-exchange contribution as well as the short-range D-term are
 important at both energies. 
 The contribution of the E term is negligible  at $E=70$~MeV but clearly
 visible at $190$~MeV.

%


In summary, we presented a new powerful calculational scheme  which enables us to take 
efficiently into account any number of contact terms of a chiral 3NF in the
3N continuum Faddeev calculations. 
 That  approach facilitates a reduction of 
 the time required to compute observables for a given set of strengths
to seconds and is thus  especially suited to repeated calculations with varying
 strengths of  contact terms. 
 We demonstrated that the proposed emulator reproduces very 
 well the exact predictions for 3N continuum  observables. 
It is conceivable that with the help of the constructed emulator of
 the exact solutions of the 3N Faddeev equation  fine
 tuning of a 3N Hamiltonian parameters based on available 3N scattering data
 is feasible.

This study has been performed within Low Energy Nuclear Physics
International Collaboration (LENPIC) project and 
was  supported by the Polish National Science Center 
 under Grant No. 2016/22/M/ST2/00173. 
 The numerical calculations were performed on the 
 supercomputer cluster of the JSC, J\"ulich, Germany.

%
%
%

\end{document}